\documentclass{aastex631}

\usepackage{gensymb}

\shorttitle{CMEs Driven by Homologous Compact-flare Blowout Jets}
\shortauthors{B. D. Patel et al.}

\graphicspath{{./}{figures/}}

\begin{document}

\title{Source Region and Launch Characteristics of Magnetic-arch-blowout Solar Coronal Mass Ejections Driven by Homologous Compact-flare Blowout Jets}

\correspondingauthor{Binal D. Patel}
\email{binalp@prl.res.in}

\author[0000-0001-5582-1170]{Binal D. Patel}
\affiliation{Udaipur Solar Observatory, Physical Research Laboratory, Dewali, Badi Road, Udaipur-313 001, Rajasthan, India}
\affiliation{Discipline of Physics, Indian Institute of Technology Gandhinagar, Palaj, Gandhinagar-382 355, Gujarat, India}

\author[0000-0001-5042-2170]{Bhuwan Joshi}
\affiliation{Udaipur Solar Observatory, Physical Research Laboratory, Dewali, Badi Road, Udaipur-313 001, Rajasthan, India}

\author[0000-0003-1281-897X]{Alphonse C. Sterling}
\affiliation{NASA/Marshall Space Flight Center, Huntsville, AL 35812, USA}

\author[0000-0002-5691-6152]{Ronald L. Moore} 
\affiliation{NASA/Marshall Space Flight Center, Huntsville, AL 35812, USA}
\affiliation{Center for Space Plasma and Aeronomic Research, University of Alabama in Huntsville, Huntsville, AL 35805, USA}

\begin{abstract} 
We study the formation of four coronal mass ejections (CMEs) originating from homologous blowout jets. All of the blowout jets originated from NOAA active region (AR) 11515 on 2012 July 2, within a time interval of $\approx$14 hr. All of the CMEs were wide (angular widths $\approx$95--150$\arcdeg$), and propagated with speeds ranging between $\approx$300--500 km s$^{-1}$ in LASCO coronagraph images. Observations at various EUV wavelengths in Solar Dynamics Observatory/Atmospheric Imaging Assembly images reveal that in all the cases, the source region of the jets lies at the boundary of the leading part of AR 11515 that hosts a small filament before each event. Coronal magnetic field modeling based on nonlinear force free extrapolations indicate that in each case the filament is contained inside of a magnetic flux rope that remains constrained by overlying compact loops. The southern footpoint of each filament is rooted in the negative polarity region where the eruption onsets occur. This negative-polarity region undergoes continuous flux changes, including emergence and cancellation with opposite polarity in the vicinity of the flux rope, and the EUV images reveal brightening episodes near the filament's southeastern footpoint before each eruption. Therefore, these flux changes are likely the cause of the subsequent eruptions. These four homologous eruptions originate near adjacent feet of two large-scale loop systems connecting from that positive-polarity part of the AR to two remote negative-polarity regions, and result in large-scale consequences in the solar corona.  
\end{abstract}
\keywords{Unified Astronomy Thesaurus concepts: Solar active regions (1974); Solar magnetic fields(1503); Solar active region filaments (1977); Solar coronal transients (312); Solar flares (1496); Solar coronal mass ejections (310)}

\section{Introduction} \label{sec:intro}
Solar eruptions that produce flares and coronal mass ejections (CMEs) are among the most spectacular, large-scale phenomena in the solar atmosphere and heliosphere. A solar flare is essentially characterized by the catastrophic and explosive energy release of the order of 10$^{27}$--10$^{32}$ ergs within tens of minutes in localized regions of the solar atmosphere. The successful ejection by the eruption of the coronal plasma and magnetic field into the heliosphere is observed as a CME, propagating outwards in the solar atmosphere and eventually entering the interplanetary medium. When an interplanetary CME hits the Earth's magnetosphere, it can produce severe space weather effects \citep[for e.g.,][]{1990GMS....58..343G, 2017LRSP...14....5K, 2021LRSP...18....4T, 2018SoPh..293..107J, 2022SoPh..297..139P}. White-light coronagraph observations of CMEs have revealed that they often show a classic ``three-part'' structure consisting of a bright front, a dark cavity, and a bright core \citep{1983JGR....8810210I}. The bright front originates from the plasma pile-up at the front of the expanding CME \citep[e.g.,][]{2003ApJ...598.1392V, 2017PhPl...24i0501C}, the cavity is interpreted as a coherent magnetic flux rope (MFR) structure \citep{2008ApJ...672.1221R}, and the bright core is the dense plasma of the expelled filament/prominence \citep{2013SoPh..284..179V}. Some of the contemporary observations challenge the traditional interpretation of the CME's substructures -- cavity and core. A survey of CMEs, in the form of four exhibits, conducted by \cite{2017ApJ...834...86H} showed that most of the CMEs lack an eruptive filament while for other events no geometrical resemblance was found between CME's bright core and corresponding eruptive filament. The observations of \cite{2023ApJ...942...19S} suggested that the dark cavity may not represent an MFR; instead, it is a low-density zone between the CME front and a trailing MFR. While above studies constraint our traditional understanding of the CME's three-part structure, there is unanimity on three-part structure being an intrinsic characteristic of CMEs, and that it has fundamental importance for understanding their origin and evolution.

The studies concerning the onset and triggering of solar flare, filament/prominence eruption, and CME have established that these are different manifestations of the same explosive magnetic eruption in the solar atmosphere \citep{2000JGR...10523153F, 2001ApJ...552..833M, 2017PhPl...24i0501C, 2018SSRv..214...46G, 2019RSPTA.37780094G}. In addition to these phenomena, a shock can also be considered as a fourth aspect of the eruption, present at the front and flanks of the CME, if the CME's speed exceeds the local Alfv\'en speed \citep{2008A&A...491..873C, 2012ApJ...744...72G, 2021SoPh..296..142P}. Contemporary observations also reveal a few special categories of CMEs, for example, narrow \citep{2001ApJ...550.1093G, 2011LRSP....8....1C}, streamer-puff \citep{2005ApJ...635L.189B, 2007ApJ...661..543M, 2016ApJ...822L..23P}, and stealth CMEs \citep{2013SoPh..285..269H, 2021ApJ...908...89O}. The understanding of such ``non-standard'' events requires detailed case studies.

The kinematic evolution of CMEs usually comprises three phases: a slow-rise phase, an impulsive main-acceleration phase, and a propagation phase with slowly varying velocity \citep{2001ApJ...559..452Z, 2004ApJ...604..420Z, 2006ApJ...649.1100Z}. The early signatures of a CME in the lower corona are various pre-eruptive features. Common such pre-eruptive structures include filaments/prominences \citep{1988assu.book.....Z, 2016ApJ...832..130J}, coronal sigmoids \citep{1996ApJ...464L.199R, 1996ApJ...468L..73M, 1997ApJ...491L..55S, 2017ApJ...834...42J, 2018ApJ...869...69M, 2020ApJ...900...23M, 2021SoPh..296...85J}, coronal hot channels \citep{2012NatCo...3..747Z, 2016ApJ...830...80K, 2019ApJ...884...46M, 2020ApJ...897..157S, 2020A&A...642A.109N}, and coronal cavities \citep{2013SoPh..288..603F, 2015ASSL..415..323G}. These pre-eruption signatures are often assumed to be aspects of a magnetic flux rope, and the term ``magnetic flux rope'' or ``flux rope'' is often used interchangeably with them \citep{2017PhPl...24i0501C, 2019RSPTA.37780094G, 2020RAA....20..165L, 2018ApJ...869...69M, 2020SoPh..295...29M, 2020ApJ...897..157S}. A flux rope is defined as a set of magnetic field lines that wrap around each other in a braided manner or wrap around a central axis \citep{2016A&A...591A..16P}. 

Previous studies support that solar eruptions can be triggered by magnetic reconnection. In this context, two models -- tether cutting and breakout-- have gained much attention. These two models explore the role of the initial magnetic reconnection in two different ways in setting up favorable conditions for the flux rope eruption. The ``tether-cutting model'' is based on a highly sheared bipolar magnetic field, with the earliest reconnection occurring deep in the sheared core (inner part of the bipole) field \citep{1992LNP...399...69M, 2001ApJ...552..833M}. In contrast, the ``breakout model'' consists of a quadrupolar magnetic topology. In this case, the eruption is initiated by reconnection at a null point lying elevated in the corona between the central sheared field and an overlying field connecting the outer polarities in the quadrupolar magnetic configuration \citep{1999ApJ...510..485A, 2007SoPh..242..143J, 2012ApJ...760...81K, 2021MNRAS.501.4703J}.

Successive flares occurring at the same locations in an active region and showing a strikingly similar pattern of structure and development are known as ``homologous flares'' \citep[e.g., ][]{MARTRES19845, 1989SoPh..119..357M}. The CMEs accompanying such homologous flares are known as ``homologous CMEs'' and display similar morphological appearance in the coronagraph observations \citep{2013ApJ...778L..29L, 2017ApJ...844..141L, 2017SoPh..292...64L, 2022ApJ...930...41S}. Previous studies have reported possible causes for homologous flare-CMEs could be continuous flux emergence in the form of moving magnetic features \citep{2002ApJ...566L.117Z}; significant disturbance, partial eruption, and relatively fast restoration of the same large-scale structures involved in the repeating CME events \citep{2004JGRA..109.2112C}; and/or shearing motion and magnetic flux cancellation of opposite-polarity fluxes at the polarity inversion line \citep{2020ApJ...901...40D}. To explore the development of a homologous sequence of three CMEs \cite{2013ApJ...778L...8C} conducted a magnetohydrodynamic (MHD) simulation and demonstrated that the reformation of a twisted flux rope after each CME eruption during the sustained flux emergence can explain the repeated reformations of low-coronal source of homologous CMEs in the form of coronal sigmoids and ``sigmoid-under-cusp'' configurations.
%\textbf{Magnetohydrodynamic simulations also indicate that the recurring occurrences of CME events originate from the iterative creation and ejection of kink-unstable flux ropes. This phenomenon arises as a consequence of the ongoing emergence of twisted flux ropes into pre-existing potential coronal arcades \citep{2013ApJ...778L...8C}.} Hence, close examination of the source regions of homologous CMEs and underlying magnetic conditions is essential for understanding the onset and evolution of the CMEs.

%Magnetohydrodynamic simulations also suggest that the repetitive CME activities originate from repeated formations and eruptions of kink-unstable flux ropes due to the result of continued emergence of twisted flux ropes into pre-existing coronal potential arcades \citep{2013ApJ...778L...8C}. 

Coronal jets appear as transient (lasting tens of minutes) coronal-temperature ejections from near the solar surface into the corona, with spires that grow to typically several times 10$^4$ km. \cite{2010ApJ...720..757M, 2013ApJ...769..134M} noted a dichotomy in the morphology of solar coronal jets, consisting of two broad categories, viz., standard and blowout jets. Each jet spire shoots out from a base that is bright in coronal emissions. Blowout jets are characterized by spires that have widths comparable to that of their bright base. Standard jets have spires that remain narrow, substantially less wide than their base. \cite{2016ApJ...833..150L} presented a detailed analysis of recurrent homologous jets originating from an emerging magnetic flux at the edge of an active region that include the cases of both standard and blowout categories. Their results underline a direct involvement of magnetic reconnection in transporting large amounts of free magnetic energy into jets and that the blowout jets release more free energy than the standard jets. \cite{2010A&A...512L...2A} investigated the origin of jets as a result of repeated reconnection events between colliding magnetic fields by performing 3D MHD simulations. Their simulations suggested the jet-like emission to be a consequence of emergence of new magnetic flux, which introduces a perturbation to the active region field, thus making favorable conditions for the magnetic reconnection between the adjacent field lines. High resolution observations contain evidence that both standard and blowout jets result from minifilament eruptions \citep{2015Natur.523..437S, 2016ApJ...821..100S, 2016ApJ...832L...7P, 2017ApJ...844...28S}. In blowout jets the erupting minifilament tends to be more explosive than in standard jets so that it fully erupts out of the base region into open coronal field; for standard jets the minifilament eruptions tend to be more confined to the base \citep{2015Natur.523..437S, 2022ApJ...927..127S}. There are further indications that the minifilament eruptions often result from magnetic flux cancellation \citep[e.g., ][]{2016ApJ...832L...7P, 2017ApJ...844..131P, 2017ApJ...844...28S, 2018ApJ...853..189P, 2019ApJ...882...16M, 2021ApJ...909..133M}. In some cases magnetic shearing or rotational motions may be important \citep{2019ApJ...873...93K, 2022ApJ...932L...9K}.

With the identification of blowout jets, special attention has been given to recognize the physical mechanism governing the jet-CME relationship. \cite{2012ApJ...745..164S}, reported an intriguing blowout jet exhibiting cool and hot components next to each other that was eventually associated with simultaneous bubble-like and jet-like CMEs. Based on their multi-wavelength and multi-angle observations, they proposed a model for the observed blowout jet that involve both external and internal reconnections -- the external reconnection produces the jet-like CME and causes the rise of the filament, while subsequent internal reconnection begins underneath the rising filament and correlates with the bubble-like CME. \cite{2015ApJ...813..115L} observed a jet-triggered CME and, based on temporal and spatial relationship between the two events, suggested the possibility for the jet to evolve as the core of the CME. \cite{2016ApJ...822L..23P} studied seven homologous jet-driven ``streamer-puff CMEs," along with non-CME-producing jets, from an active region. They found that the CME-producing jets were blowout jets and expelled cool plasma material with higher speed than the non-CME-producing jets. \cite{2016ApJ...819L..18Z} studied an interesting case of interaction of jet and neighboring coronal hole (CH) that resulted into the noticeable changes in the morphology of the jet front. Their study reveals that the jet-CH collision and the large-scale magnetic topology of the CH play important role in defining the eventual propagation direction of this jet-CME eruption. To explore the link between jet-producing minifilament/flux-rope eruptions and CME-producing filament/flux-rope eruptions from ARs, \cite{2018ApJ...864...68S} tracked two small, magnetically isolated ARs for several days, from the time of their emergence on the solar disk until a time when the regions evolved into the source of CME-producing eruptions. Their observations confirm that the erupting flux ropes form at sites of flux cancellation \cite[see also ][]{2011A&A...526A...2G, 2017ApJ...844..131P} and that the minifilaments are small-scale versions of the long-studied full-sized filaments.

In this paper, we analyze four homologous large-scale blowout jets that produced spectacular CMEs in coronagraphic images. The eruptions originated from NOAA active region (AR) 11515 within a period of $\approx$14 hours on 2012 July 2. Notably, each event originated from the same part of the AR. Several aspects of our events I and III (i.e., the events with flare of classes M5.6 and M3.8, respectively, in Table~\ref{tab:hom_cme}, discussed below) have been reported in previous studies that utilized space and ground-based observations \citep{2014A&A...562A.110L, 2016ApJ...817...39L, 2021ApJ...911...33C}. The scope of our study is much broader, in that here we find evidence of a common physical mechanism for the production of comparatively wide CMEs that emanate from four comparatively compact homologous blowout-eruption flares. Further, the focus of our study is a detailed investigation of the large-scale coronal magnetic fields stemming from the flaring region and their roles in the development of the CMEs' structure. All four CMEs are produced by blowout eruptions of relatively small-scale filaments, which eruptions make large-scale coronal dimming. Section~\ref{sec:obs_data} provides information on the observational data sources and techniques. The details of the analysis and results are given in Section~\ref{sec:analysis_results}. The discussions and interpretation of our study are provided in Section~\ref{sec:discussion}.

\section{Observational Data and Analysis Techniques} \label{sec:obs_data}

We have utilized data obtained from the Atmospheric Imaging Assembly \cite[AIA;][]{2012SoPh..275...17L} on board the Solar Dynamics Observatory \cite[SDO;][]{2012SoPh..275....3P} for extreme ultraviolet (EUV) imaging and analysis. AIA observes the Sun in seven different EUV wavelengths, including 304 \AA, 171 \AA, 193 \AA, 211 \AA, 335 \AA, 94 \AA, and 131 \AA. Additionally, the AIA images the solar atmosphere in two UV wavelengths at 1700 \AA~and 1600 \AA~and in one visible wavelength at 4500 \AA. The AIA images have a temporal cadence of 12 s and 24 s for EUV and UV passbands, respectively, with a pixel resolution of 0$\farcs$6. For the present study, we have utilized the 12 s cadence, full disk observations from the AIA in various EUV channels. For chromospheric images of the Sun, we have used 2048 $\times$ 2048 pixel full disk images in the H$\alpha$ channel with a pixel resolution of $\approx$1$\farcs$0, obtained from the archival data of global oscillation network group \cite[GONG;][]{1996Sci...272.1284H}. 

In this study, we investigated photospheric structures of AR 11515. For this purpose, we used 4096 $\times$ 4096-pixel full disk line-of-sight (LOS) continuum, and magnetogram observations recorded by the Helioseismic and Magnetic Imager \cite[HMI;][]{2012SoPh..275..229S} on board SDO. The HMI magnetograms have a temporal cadence of 45 s and spatial resolution of 0$\farcs$5 pixel$^{-1}$. To compare the HMI data with AIA data, we employ the routine \textit{hmi\_prep.pro} to convert the resolution of HMI 0$\farcs$5 pixel$^{-1}$ into 0$\farcs$6 pixel$^{-1}$ in the SolarSoftWare package \citep{2012ascl.soft08013F}. 

We have used images from the X-ray Telescope (XRT) on board Hinode to view the AR in soft X-ray (SXR) emission \citep{2007SoPh..243...63G}. Hinode/XRT provides a broadband response in the soft X-ray range to emissions from plasma hotter than about 1 MK. XRT has a spatial resolution of 1$\farcs$02 pixels, and usually observes the Sun with a limited field of view with varying cadences. 
 
To deduce the configuration of coronal magnetic fields in the corona before flares, we employ the non-linear force-free field (NLFFF) extrapolation method developed by \cite{2004SoPh..219...87W, 2008JGRA..113.3S02W}, which is based on optimization techniques. For this, we used vector magnetogram data from the hmi.sharp\_cea\_720s series of HMI. We convert the magentograms to 1$\farcs$0 pixel$^{-1}$ to provide the input boundary condition for the extrapolation. Our extrapolation volume spans 576, 232, and 192 pixels in the x, y, and z directions, respectively. These correspond to a physical volume of 418 $\times$ 168 $\times$ 139 Mm$^3$ above the solar surface of the AR under consideration. For visualization of the modeled extrapolated field lines in the AR volume, we have used the Visualization and Analysis Platform for Ocean, Atmosphere, and Solar Researchers software \citep{2007NJPh....9..301C}. 

The CME events under analysis were observed by the Large Angle and Spectrometric Coronagraph \citep[LASCO;][]{1995SoPh..162..357B} C2 and C3 coronagraphs on board the Solar and Heliospheric Observatory \citep[SOHO;][]{1995SoPh..162....1D}. The C2 and C3 coronagraphs observe the solar corona in the white light with 1.5--6 and 5.5--30 R$_\sun$ field of view (FOV), respectively. We also utilize the LASCO/CDAW catalog\footnote{\url{https://cdaw.gsfc.nasa.gov/CME\_list/}} to obtain CME parameters such as linear speed, acceleration, and angular width. 

We found a total of four CME-producing homologous compact-flare blowout coronal jet eruptions from National Oceanic and Atmospheric Administration (NOAA) AR 11515 over a $\approx$14 hour-long interval on 2012 July 2. These CMEs and their source eruptions were identified by looking at movies from the Solar and Heliospheric image Visualization tool: Helioviewer\footnote{\url{https://helioviewer.org/}}. Table~\ref{tab:hom_cme} lists the events under consideration for the present study, which are four blowout-jet-making filament eruptions and their corresponding CMEs and accompanying compact flares.

\section{Analysis and Results} \label{sec:analysis_results}

\subsection{Homologous Coronal Mass Ejections and their Source Region}
\begin{figure}[ht!]
\plotone{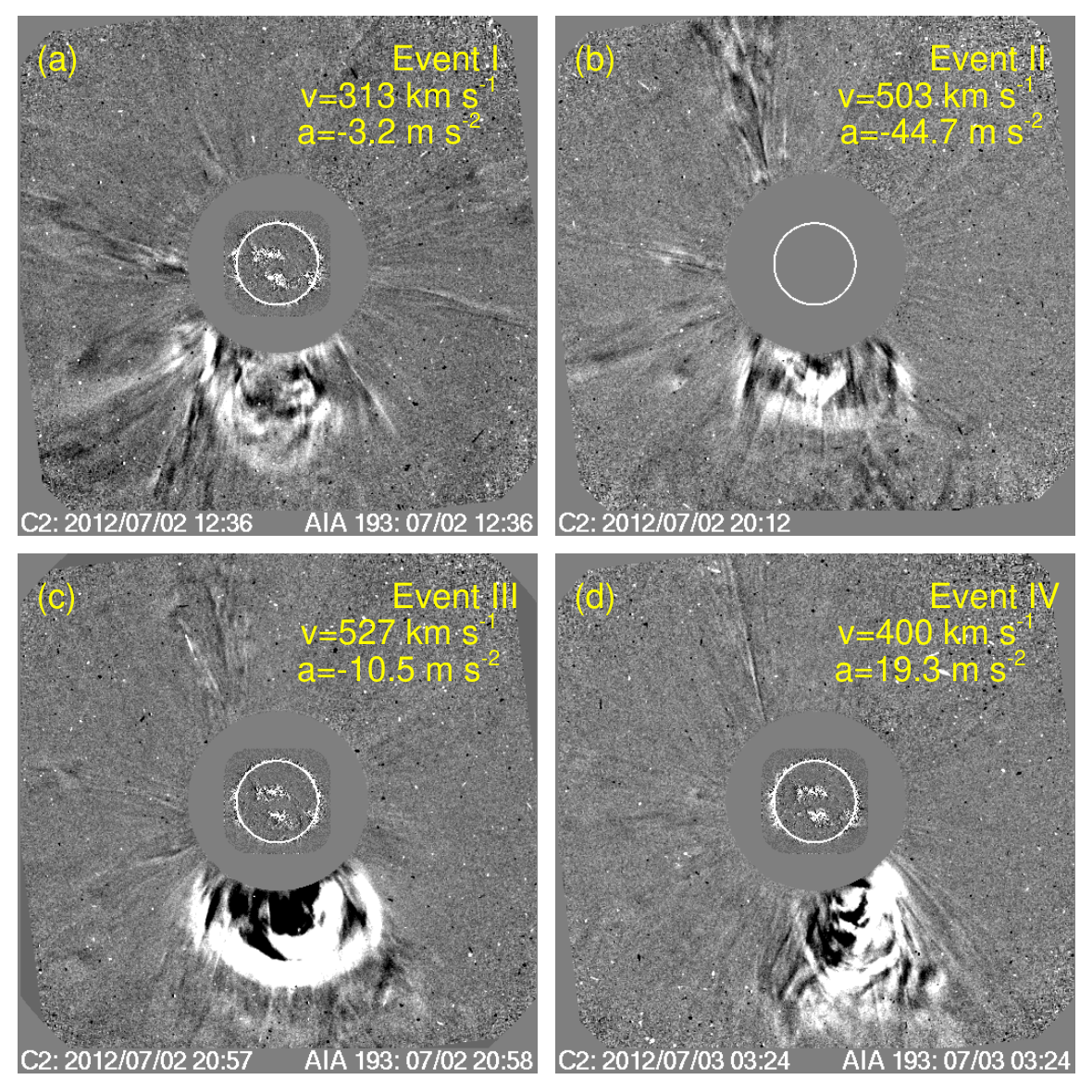}
\caption{Representative images of CMEs observed from the LASCO C2 coronagraph depicting the homologous CMEs from NOAA AR 11515. Panels (a), (b), (c), and (d) show the CMEs in events I, II, III, and IV, respectively. The CME speed and acceleration from the LASCO CME catalog are annotated in each panel.
\label{fig:cme}}
\end{figure}

In Figure~\ref{fig:cme}, we provide LASCO C2 images of the four homologous CMEs of events I, II, III, and IV of Table~\ref{tab:hom_cme}. The CMEs of events I and III are partial-halo CMEs (with angular sizes of 125\degr~and 145\degr, respectively). In contrast, the CMEs of events II and IV have comparatively narrow angular widths of 96\degr~and 97\degr, respectively. The linear speed of these CMEs ranges between $\approx$ 300--530 km s$^{-1}$ (see Table~\ref{tab:hom_cme}). 

In Figure~\ref{fig:overview}, we present a multiwavelength overview of NOAA AR 11515. In panels (a)--(d), we show simultaneous white-light continuum, magnetogram, and EUV images before the onset of the event I (M5.6 GOES flare class). Panels (a) and (b) show there were two adjacent ARs (i.e., ARs 11515 and 11514), enclosed in the two black boxes in panels (a)--(d). The leading part of the AR 11515, which predominantly contains two positive polarity sunspot groups (P$_1$ and P$_2$), hosts the source of the eruptive flares under analysis. The flaring region is inside the green box in panel (b). There is emerging negative polarity flux (labeled N$_E$) in the flaring region. The trailing part of the AR contains small sunspots predominantly of negative polarity flux (labeled N$_1$). In HMI magnetograms (Figure~\ref{fig:overview}b), there is a distant negative polarity region (labeled N$_2$) located to the southwest of AR 11515. The EUV images in the 171 and 211 \AA~channels (see Figures~\ref{fig:overview}c and d) show large coronal loops connecting the leading and trailing parts of the AR 11515. These EUV images further show large-scale loops that connect the two ARs. EUV images from some channels (e.g., 211 \AA~and 193 \AA) show large-scale loops connecting the P$_1$ region of AR 11515 with the N$_2$ region. In Figure~\ref{fig:hinode_xrt}, we provide an Hinode-XRT image of our region of interest. Here, we clearly recognize hot, large-scale coronal loops connecting P$_1$--N$_1$ and P$_1$--N$_2$ regions, which we identify as LSL1 and LSL2, respectively. The existence of LSL1 and LSL2 and its role in the backdrop of the homologous eruptions are elaborated and further clarified in Section~\ref{sec:onset_dimming} (see Figure~\ref{fig:extrapolation_pfss}).

To examine the configuration of the flaring region before each event, we present images of the flaring region in AIA 304 \AA~and GONG H$\alpha$ in Figures~\ref{fig:overview}(e1)-(e4) and (f1)-(f4), respectively. In both AIA 304 and GONG H$\alpha$ images, we identify a filament prior to each jet-flare event, which implies that either the filament did not totally erupt in each of the events, or the filaments recursively formed in the filament channel between the positive polarity sunspot (P$_2$) and the negative polarity flux (N$_E$) next to P$_1$. The filaments in the AIA 304 \AA~channel and H$\alpha$ filtergrams present similar morphological structure. In Section~\ref{subsec:extrapolation} (see Figure~\ref{fig:extrapolation_fr}(a3)--(d3)), we further confirm the existence of an MFR associated with the filament.

%Both AIA 304 and GONG H$\alpha$ images show \textbf{filaments} before each flare, which implies that either the filament did not totally erupt in each of the events, or the filaments recursively formed in the filament channel between the positive polarity sunspot (P$_2$) and the negative polarity flux (N$_E$) next to P$_1$. The filaments in the AIA 304 \AA~channel and H$\alpha$ filtergrams present similar morphological structure. 

Figure~\ref{fig:goes_ew_overview} presents an overview of the temporal and spatial evolution of the eruptive flares. In Figure~\ref{fig:goes_ew_overview}(a), we show the GOES light curve in the 1--8 \AA~band and AIA light curves in various channels (94 \AA, 171 \AA, 304 \AA) from 09:00:00 UT on 2012 July 2 to 01:00:00 UT on 2012 July 3. We mark the occurrence of the four eruptive flares of class M5.6, C7.4, M3.8, and M2.0 in our four events (Table~\ref{tab:hom_cme}) with the yellow shaded bands. The GOES lightcurve indicates that flare I of GOES M5.6 class began at $\approx$10:43 UT. This short-lived flare reached its peak at 10:52 UT. Flare II of GOES C7.4 class began at $\approx$18:45 UT and peaked at $\approx$18:56 UT. Similarly, the subsequent two flares that originated from this region were of similarly short duration. That is, all four flares had GOES durations of $\approx$14 to 17 minutes.

In Figure~\ref{fig:goes_ew_overview}(b)--(m), we present a few representative images of each flare in the AIA 94 [log(T)=6.8], 171 [log(T)=5.7], and 304 [log(T)=4.7] \AA~channels. In Figures~\ref{fig:goes_ew_overview}(b)--(e), we present AIA images from the 94 \AA~channel near the peak time of the flares. Similarly, panels (f)--(i) and (j)--(m) show EUV images near the peak time of the flares in 171 and 304 \AA, respectively. The AIA images show that the eruptions were compact and produced collimated eruptions (see also the animation accompanying with Figure~\ref{fig:goes_ew_overview}). Furthermore, these eruptions occur from the same location in the AR within $\approx$14-hour period and presented similar morphological evolution, showing the four eruptions to be homologous in nature. 

The filament before each flare lies below compact closed field lines, and the eruption becomes a collimated jet-like structure. The imaging observations reveal the morphology of the eruptions at the solar source region to be an inverted Y (or inverted lambda)-shaped structure, basically the blowout jet morphology \citep{2010ApJ...720..757M}. Blowout jets observed previously over polar coronal holes, quiet regions, and near ARs are smaller-scale eruptions. We do not discuss jets and blowout jets in detail here, but see, e.g., \cite{2015Natur.523..437S, 2016ApJ...821..100S, 2016ApJ...832L...7P, 2017ApJ...844...28S, 2022ApJ...927..127S}. In our case, the eruptions occurred in the AR, are much larger, and produced still-larger-scale consequences.

\subsection{Eruption Onset and Coronal Dimming}
\label{sec:onset_dimming}

For quantitative tracking of the eruptions, we present time-slice diagrams in Figure~\ref{fig:ht_merge}, obtained from the plane of sky projected images. Figures~\ref{fig:ht_merge}(a)--(b), (c)--(d), (e)--(f), and (g)--(h) correspond to events I, II, III, and IV, respectively. In the left column, we show the slices along which we track the erupting structures in AIA 304 \AA~images. In the right column, we present the corresponding time-slice diagram. To compare the kinematic evolution of erupting material, which is comprised of both cool filamentary material and hot plasma, with the temporal evolution of each flare, we overplot the AIA 304 \AA~flare lightcurve on the time-slice diagram. Each event produced a CME contain substantial plasma erupted from the lower solar atmosphere (i.e., chromosphere, transition region, and lower corona). We find that, except for event II, the rise of the flare intensity is nearly simultaneous to the start of fast eruption of the filament. We also annotate the speed of the ejecting plasma in the time-slice diagrams shown in panels (b), (d), (f), and (h). Leading to event II, jets were occurring continuously near the south-eastern footpoint of the filament, which produced episodic plasma eruptions and also gave rise to intensity enhancements in the AIA 304 \AA~flux profile. These jets are followed by CME-producing larger eruption (traced by the dotted line in panel d). Notably, event III shows a slow-rise phase during the pre-flare phase prior to its fast-flare phase, consistent with an earlier study of this event \citep{2016ApJ...817...39L}.

\begin{figure}[ht!]
\epsscale{0.9}
\plotone{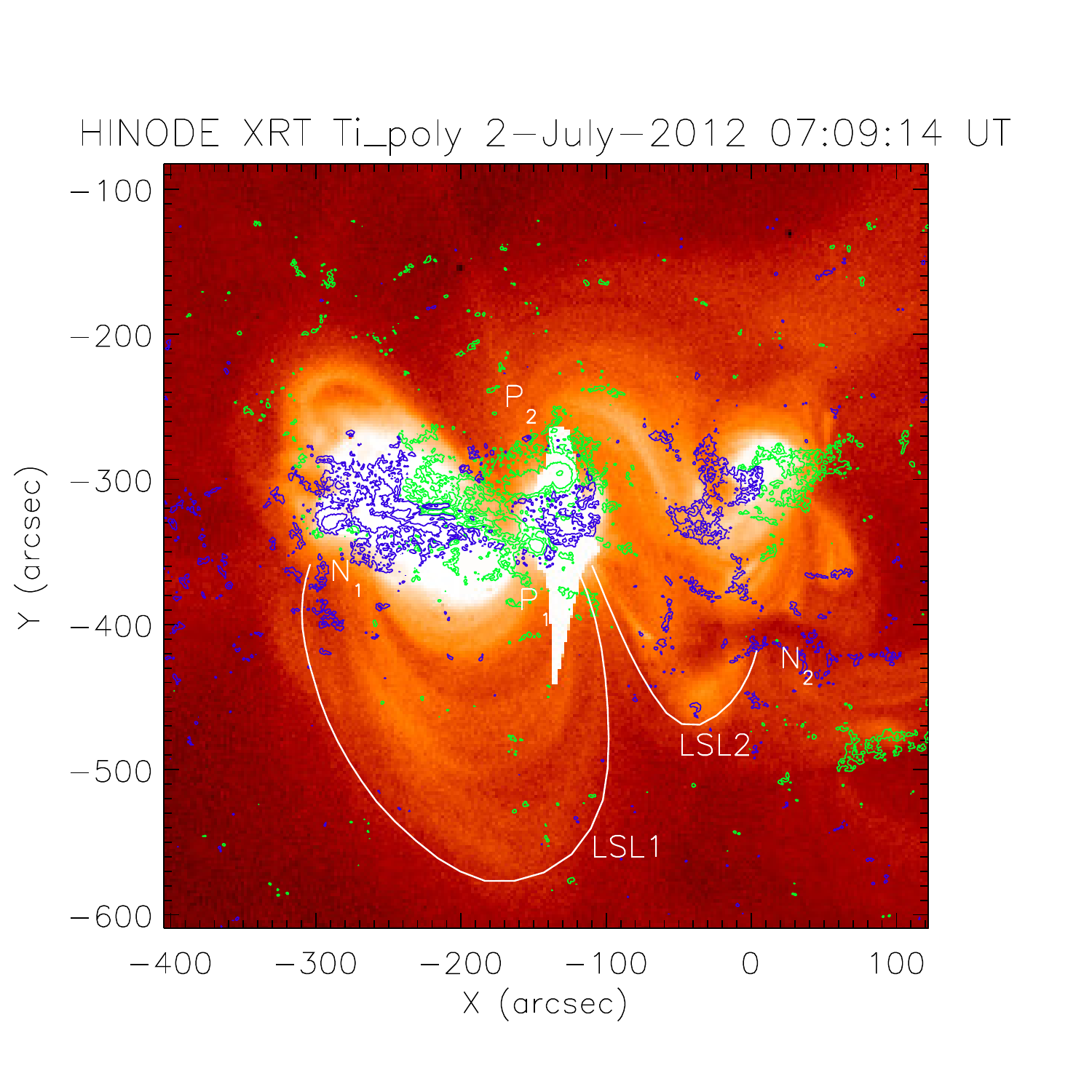}
\caption{The HINODE X-ray Telescope image of the AR 11515 on 2012 July 2 at 07:09 UT, showing large-scale loops connecting remote opposite-polarity magnetic flux, P$_1$ and N$_1$ (LSL1), and P$_1$ and N$_2$ (LSL2).
\label{fig:hinode_xrt}}
\end{figure}

Each of the flares erupted in the same compact part of the active region, and produced a collimated blowout jet-like eruption; however, their CMEs have broad angular widths ($\sim$96--145\degr). Therefore, we examine the large-scale coronal changes resulting from the blowout jets to assess why the CMEs are so wide. To display the large-scale consequences of the blowout jets, we show AIA 193 \AA~fixed-difference images in Figures~\ref{fig:dimming_mosaic}. We mark the flaring region by the green box in Figure~\ref{fig:dimming_mosaic}(a) to provide a comparison between the areas of the flaring and dimming regions. We find that the eruption of the filament from the flaring region is followed by EUV dimming for all the four cases (see the animation accompanying Figure~\ref{fig:dimming_mosaic}). In all the events, the coronal dimming is disjoint from the location of the filament and flare. Notably, the main dimming region is located southwest of the jet base and undergoes considerable expansion; the dimming area is shown inside a blue ellipse. As each eruption progresses, there is successive brightening and darkening (``glittering") at the trailing part of AR 11515 (see the region inside the cyan circles in panels (a)--(d)). Contextually, the glittering region is the footpoints of large-scale loops that connect the leading and trailing parts of the AR (i.e., LSL1, see Figure~\ref{fig:hinode_xrt}). Hence, we term this feature as ``footpoint (FP) glittering.'' The region of coronal dimming spatially corresponds to the dispersed negative magnetic polarity region (N$_2$), which is connected to the flaring region by the large-scale loops LSL2 (see Figures~\ref{fig:overview} and \ref{fig:hinode_xrt}).

To verify the large-scale magnetic connectivity of the solar corona before the eruption, we use the Potential Field Source Surface (PFSS) extrapolation model field. In Figure~\ref{fig:extrapolation_pfss}, we show the HMI LOS magnetogram (panel a) and corresponding PFSS extrapolation model (panel b). We mark the flaring region with the green box in panel a, and we mark the location of the MFR identified in the NLFFF model by a green circle in panel b (see also Section~\ref{subsec:extrapolation} and Figure~\ref{fig:extrapolation_fr}). The PFSS extrapolation results show a large-scale connectivity between the leading and trailing parts of the active region AR 11515, marked as LSL1. The magnetogram and PFSS extrapolated magnetic field lines indicate the large-scale coronal magnetic connectivity to farther from the flaring region, with connecting field lines between positive polarities (P$_1$ and P$_2$) and dispersed negative polarity marked as N$_2$, i.e., LSL2. Interestingly, these extrapolated field lines connect to the areas where we observe coronal dimming in each of the events. To examine the magnetic field changes in the vicinity of the filaments, we discuss the magnetic flux evolution in detail in the following section.

\subsection{Structure and Evolution of the Magnetic Field}\label{sec:mag_field}
\subsubsection{Photospheric magnetic field}
AR 11515 appeared on the eastern limb of the Sun on 2012 June 28 as a simple $\beta$-type (bipolar) AR. The AR transformed into a relatively more complex $\beta\gamma$ type on 2012 June 30 and remained so for the next four days. It was larger in area on 2012 July 2 than on previous days\footnote{\url{http://helio.mssl.ucl.ac.uk/helio-vo/solar_activity/arstats/arstats_page5.php?region=11515}}. On 2012 July 2, the westernmost part (leading part) of the AR showed a complex magnetic field configuration, which we study in this section. The magnetic structure of this AR became more complex on 2012 July 4, of the $\beta\gamma\delta$-type, and this persisted until 2012 July 7. After 2012 July 8, the AR became less complicated and exhibited $\beta\gamma$-type. The AR disappeared from the western limb of the Sun on 2012 July 10 with $\beta\gamma$-type magnetic configuration.

We present the temporal and spatial evolutions of the photospheric magnetic fields of the flaring region in Figure~\ref{fig:hmi_flux}. In Figure~\ref{fig:hmi_flux}(a), we present the flux evolution time profiles for the positive and negative polarities for the region of interest (shown in panels (b) to (m)), together with the AIA 94~\AA~light curve. The time profiles of the magnetic fluxes are shown for the period of 04:40:00 UT on 2012 July 2 to 01:00:00 UT on 2012 July 3 ($\approx$20 hrs). This time duration covers our four flare events. In Figure~\ref{fig:hmi_flux}(a), we mark six different times (namely t$_1$, t$_2$, t$_3$, t$_4$, t$_5$, and t$_6$) to explore the spatial changes in the photospheric magnetic field distribution. The time instances t$_2$, t$_4$, t$_5$, and t$_6$ are selected approximately at the peaks of the four flares. Panels (b)--(m) presents HMI continuum and LOS magnetogram images at the times (t$_1$ -- t$_6$), which reveals the structural evolution of the magnetic flux in the selected region.

In Figures~\ref{fig:hmi_flux}(b)--(g), we show the evolution of the photosphere between times t$_1$, t$_2$, and t$_3$, using cotemporal continuum and magnetogram observations. These images present changes before and after event I (i.e., flare of class M5.6). The continuum images show the growth of the compact sunspot groups near P$_{1}$, marked by black arrows in panels (b)--(d). We mark the different polarities in panel (e) with blue arrows, where P$_{1}$, P$_2$, and N$_{E}$ represent the different positive polarities and emerging negative polarities, respectively (see Figure~\ref{fig:overview}). The filament lies in the region between positive polarities (P$_1$ and P$_2$) and emerging negative polarity region N$_{E}$. The cyan arrows in Figures~\ref{fig:hmi_flux}(e) and (f) indicate that the compact positive flux decreases before event I, followed by a reduction and then an increase of the positive flux at the locations marked by the green arrows (panels (e) to (g)). The red circle in panel (f) mark the southeastern footpoint of the filament prior to the first event.

We present the continuum and LOS magnetogram images showing the evolution of the flaring region during times t$_4$, t$_5$, and t$_6$ in Figure~\ref{fig:hmi_flux}(h)--(m), corresponding to the peak time of the C7.4, M3.8, and M2.0 class flares (events II, III, and IV), respectively. We mark the southeastern footpoint location of the filament with a red circle in panels (k)--(m). We observe the growth in the length of the polarity inversion line at the leftmost part of the flaring region (near P$_1$), which is marked by blue arrows in panels (g) and (k). During the interval between t$_5$ and t$_6$, we observe that the westernmost part of the southeastward moving sunspot group becomes more compact (indicated by black arrows in Figures~\ref{fig:hmi_flux}(i) and (j)). The corresponding magnetograms suggest that flux cancellation occurs in the N$_{E}$ region, which we mark by pink ellipses in panels (l) and (m). Further, panels (l) and (m) show a decrease in the negative flux area at the locations shown by the red arrows near the filament footpoint, consistent with flux cancellation occurring before event IV. This flux cancellation is also consistent with Figure~\ref{fig:hmi_flux}(a). Furthermore, panels (e)--(g) and (k)--(l) show clear signatures of negative-polarity flux emergence. The positive flux also shows emergence and cancellation at some locations in panels (e)--(g) and (k)--(m). Hence, these magnetogram observations clearly show a complex role of both flux emergence and flux cancellation in the selected region, and these complex dynamic flux changes are also reflected in the flux evolution profile in Figure~\ref{fig:hmi_flux}(a). Notably, rapid and continuous changes occur at the filament's footpoint in the emerging negative polarity prior to each event.

\subsubsection{Coronal magnetic field configuration}\label{subsec:extrapolation}
To understand the magnetic connectivities and topology of the eruption region and AR, we employ coronal magnetic field modeling using the NLFFF extrapolation technique and compare the relevant extrapolation features with the observed ones (Figure~\ref{fig:extrapolation}). For reference, we show an AIA 171~\AA~image of the active region before the event I in Figure~\ref{fig:extrapolation}(a) along with photospheric magnetic flux. Figures~\ref{fig:extrapolation}(b) and (c) demonstrate the presence of MFR before the event I (shown in blue) and event IV, respectively, and, also provide AR's magnetic field configuration within the extrapolation volume. Notably, the large-scale coronal magnetic field connectivity is similar before events I and IV, suggesting that although the flaring region's photospheric magnetic field has changed significantly, the large-scale magnetic connectivity remains mostly unaltered. The observed coronal loops in AIA 171 \AA~clearly indicate the connectivity between the leading (positive polarities P$_1$ and P$_2$) and trailing (negative polarity N$_1$) parts of AR 11515, LSL1. The NLFFF extrapolated field lines readily confirm the presence of LSL1 (refer to Figure~\ref{fig:hinode_xrt} and \ref{fig:extrapolation_pfss} showing the large-scale loop connectivities denotes as LSL1 and LSL2). Due to the restricted FOV of HMI SHARP series vector magnetogram, the extrapolated volume does not cover the distant negative polarity region (N$_2$; marked in Figures~\ref{fig:hinode_xrt} and \ref{fig:extrapolation_pfss}), thus not permitting to trace LSL2 in NLFFF modeling.

To explore the intricacies of the flux rope and its environment, we provide a zoomed view of the extrapolation results within the flaring region during the pre-flare phase for each event together with co-temporal AIA 304 \AA~image in Figure~\ref{fig:extrapolation_fr}. Panels (a1)--(d1) of Figure~\ref{fig:extrapolation_fr}, show the identified MFRs and nearby/overlying coronal loops in blue and green colors, respectively. Panels (a2)--(d2) of Figure~\ref{fig:extrapolation_fr}, show the side view of MFRs shown in panels (a1)--(d1), while panels (a3)--(d3) present the MFRs with AIA 304 \AA~image in the background. Notably, the identified MFR for each event runs along a filament observed in AIA 304 \AA~image. In our analysis, the MFRs have been identified based on high twist number distribution in the extrapolation volume \citep{2006JPhA...39.8321B}. The twist number values for the flux ropes are $-$2.4~$\pm$~0.4, $-$2.6~$\pm$~0.5, $-$2.2~$\pm$~0.4, and $-$2.3~$\pm$~0.4 before events I, II, III, and IV. In all cases, the MFR lies between positive polarity regions (P$_2$) and the emerging negative polarity flux, N$_E$, at the leading part of the AR (see also Figure~\ref{fig:overview} and \ref{fig:extrapolation_pfss}). A system of low-lying, closed field lines connects the positive and negative polarities, which envelop and constrain the flux ropes (green field lines). We also note that the pre-eruption flux ropes identified with the homologous eruptions are low lying (maximum height $\approx$ 5 Mm). As described earlier, the southern end of the flux rope lies in the emerging flux region of negative polarity (see Figure~\ref{fig:hmi_flux} and related description in Section~\ref{sec:mag_field}). Importantly, in all the cases, the eruption initiates at the southern end of the filament/MFR, anchored at N$_E$, which subsequently destabilizes the whole filament toward eruption expansion. The AIA 304 \AA~images further provide important clues toward the triggering reconnection at the chromospheric/transition region heights, allowing tether-release for the stressed flux rope, at the N$_E$ region in the form of localized pre-flare brightenings at the southern footpoint of the filament/MFR system (panels a3--d3).

\section{Discussion and Interpretation} \label{sec:discussion}
\begin{figure}[ht!]
\epsscale{0.9}
\plotone{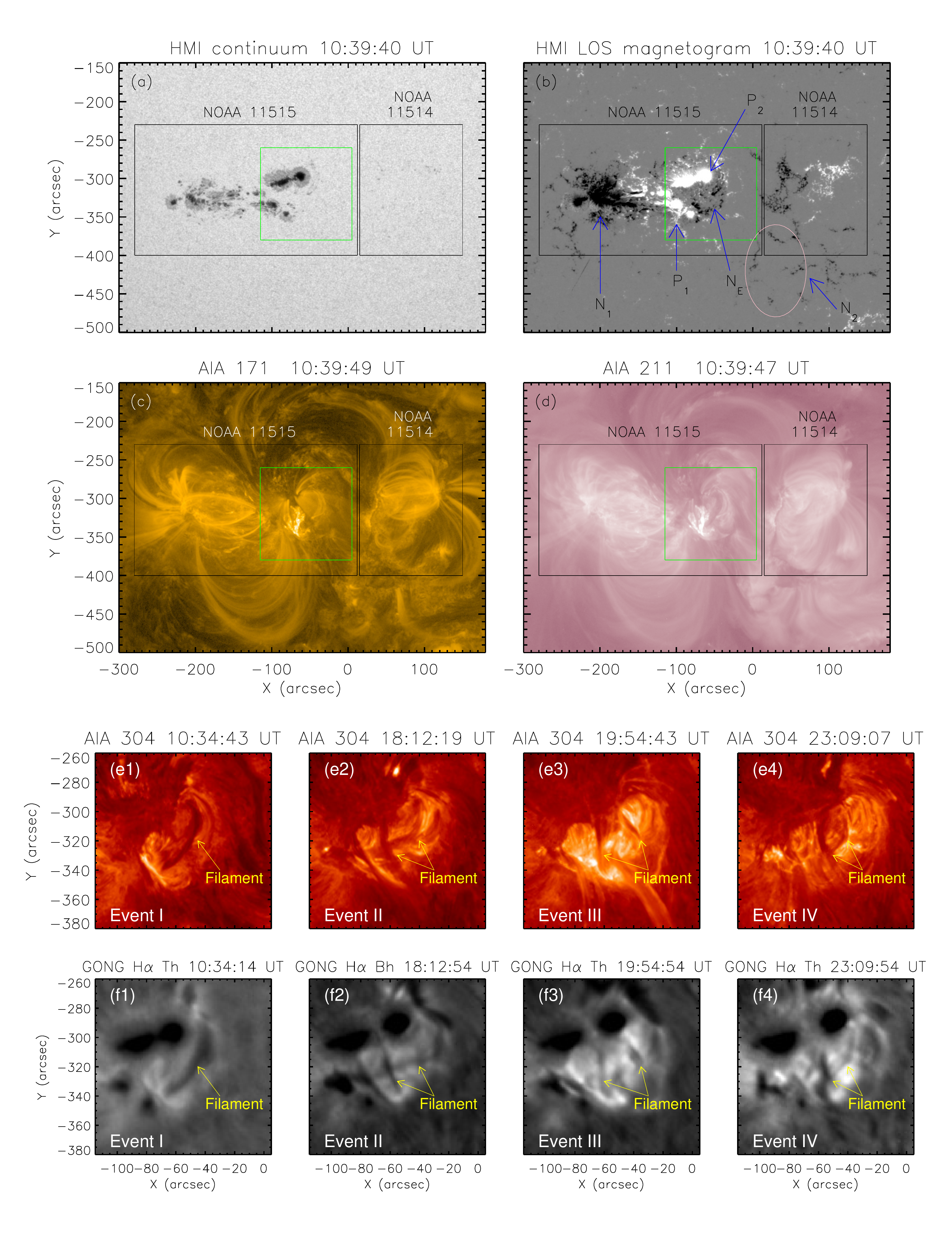}
\caption{Multiwavelength view of AR 11515 on 2012 July 2 at 10:39 UT (5 minutes before the start of the event I flare according to GOES). Panels (a) and (b) present HMI continuum and magnetogram images, respectively. We mark two ARs (ARs 11515 and 11514) with black boxes. The four homologous flares are produced by the leading part of AR 11515, which we mark with a green box. We also annotate the different magnetic polarities of the AR (N$_{1}$, P$_{1}$, N$_{E}$, and P$_{2}$) and nearby distant negative polarity region (N$_{2}$) in panel (b). Panels (c) and (d): AIA 171 \AA~[log(T)$\approx$5.7] and 211 \AA~[log(T)$\approx$6.27] images showing the connectivity of the coronal loops. Panels (e1)-(e4): AIA 304 \AA~channel images of the flaring region before each flare. Panels (f1)-(f4): Corresponding H$\alpha$ images of the flaring region, recorded by the GONG station indicated in the title of each panel as Th (Teide Observatory, Canary Islands), Bh (Big Bear Solar Observatory, USA). The yellow arrows point to the filaments before the flares.
 \label{fig:overview}}
\end{figure}

\begin{figure}[ht!]
\epsscale{0.9}
\plotone{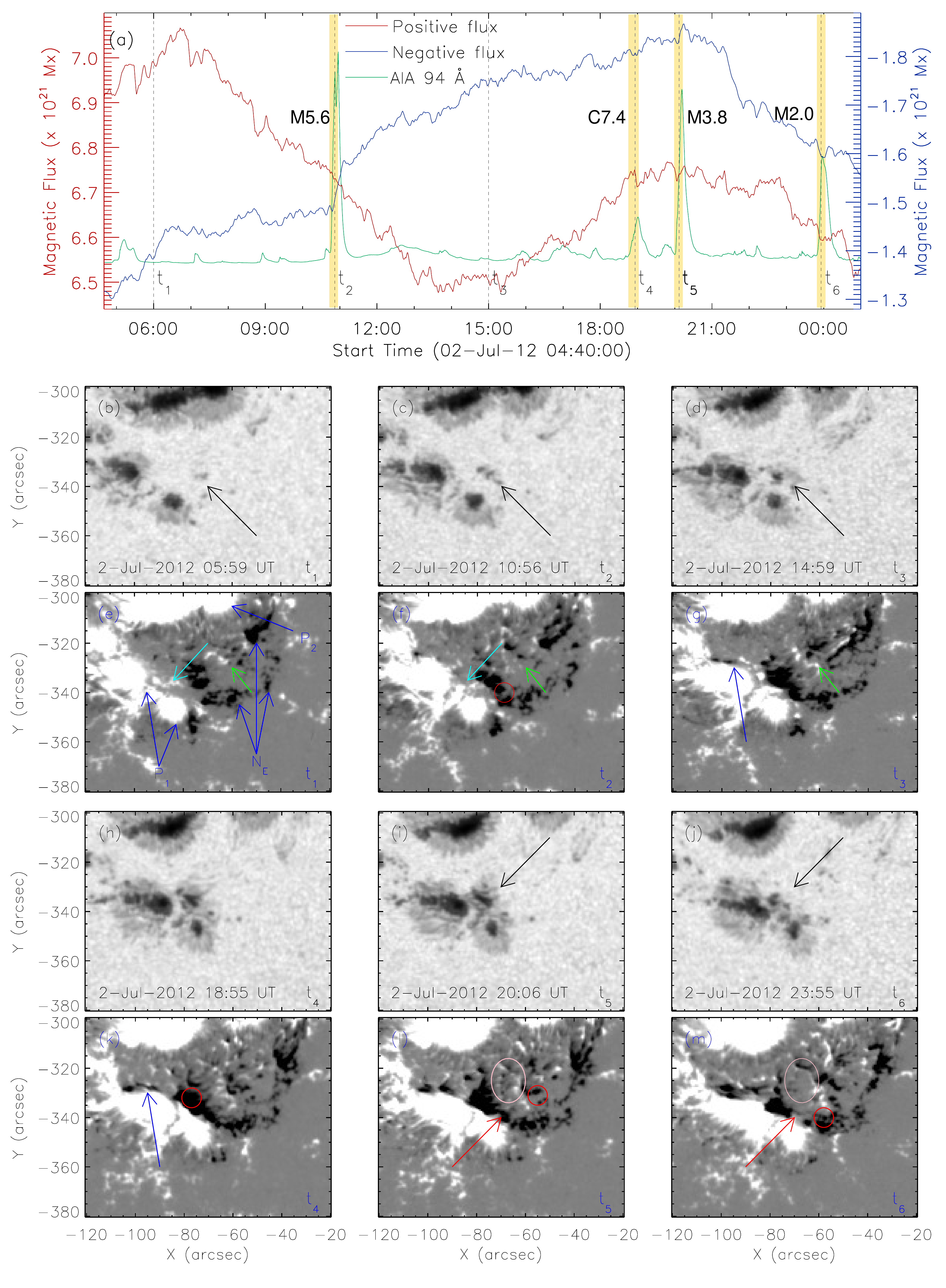}
\caption{Panel (a): The temporal evolution of magnetic flux obtained from the flaring region of the AR 11515 along with the AIA 94 \AA~channel light curve (green curve). Panels (b)--(d): HMI continuum images of the flaring region at the three times t$_1$, t$_2$, and t$_3$ marked in panel (a). The black arrows mark small-scale changes in the southwestern sunspot group. Panels (e)--(g): HMI LOS magnetogram images co-temporal with the continuum observations at times t$_1$, t$_2$, and t$_3$ (shown in the previous row). We mark evolving magnetic polarities with blue arrows in panel (e). The green arrows point to changes in the magnetogram in positive fluxes. The cyan arrows point to changing positive fluxes in panels (e) and (f), while the red circles in panels (f), (k), (l), and (m) mark the location of the southeastern footpoint of the filament. Panels (h)--(j): HMI continuum images of the flaring region at the three-time instances t$_4$, t$_5$, and t$_6$, as marked in panel (a). In Panels (i) and (j), the black arrows point to decrease in the area of the sunspot. Panels (k)--(m): HMI LOS magnetogram images at co-temporal with the continuum observations at times t$_4$, t$_5$, and t$_6$ shown in the previous row. The blue arrow in panels (g) and (k) point to changed negative flux. The red arrows in panels (l) and (m) point out a decrease in negative flux, while pink ellipses mark flux-cancellation regions.}
\label{fig:hmi_flux}
\end{figure}

\begin{figure}[ht!]
\epsscale{0.6}
\plotone{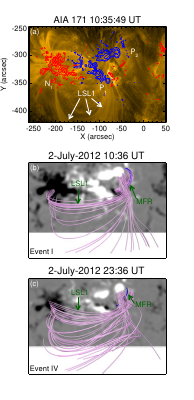}
\caption{Panel (a): AIA 171 \AA~image of the AR 11515 at $\approx$10:35 UT prior to event I, overplotted with an HMI LOS magnetogram. The positive and negative polarities of the magnetogram are shown by blue and red colors, respectively, with contour levels set as $\pm$[500,1000,1500] G. White arrows point to large-scale loops connecting leading and tailing parts of AR 11515 (i.e., LSL1). Coronal magnetic field lines obtained using the NLFFF extrapolation model are shown in panels (b) and (c) prior to events I and IV, respectively. The lower boundary of the extrapolation is the photospheric vector magnetic field. In panels (b) and (c) a green arrow points to pink field lines of large-scale loop structure (LSL1). We show the magnetic flux ropes (MFRs) in blue in all panels.
\label{fig:extrapolation}}
\end{figure}

The homologous compact blowout-jet-making filament eruptions studied in this paper are intriguing due to the following observational characteristics:
\begin{enumerate}
\item {Each flare occurred at the base of a large-scale coronal loop system that is rooted in the positive polarity region of the leading part of AR 11515, and mainly fans out toward the east (trailing part of the AR) and toward the southwest (dispersed negative polarity region outside the AR).}
\item {Each flare occurs in conjunction with the jet-like expulsion of cool plasma of an erupting small filament. Notably, in each case, the filament also manifests in NLFFF magnetic field modeling as a MFR.} 
\item{Unlike the cases of classical CME-producing solar eruptive events, the eruptions from the source regions are not wide spread, and in fact resembles a large-scale coronal blowout jet. This, however, is not unusual for AR jets, some of which have been observed to produce comparatively broad CMEs \citep{2016ApJ...822L..23P}.}
\item{The southern end of the filament is embedded in a negative-polarity region. The configuration of photospheric magnetic field at this location is extremely complex and evolves (emergence and cancellation) throughout the period of the homologous eruptions.}
\end{enumerate}
The source AR of the homologous eruptions, i.e., AR 11515, is largely bipolar with distinct leading and trailing parts (Figure~\ref{fig:overview}). The large-scale flux at the trailing part is negative and the leading part is positive; the leading part includes a less dominant but significant negative-polarity region (Figure~\ref{fig:hmi_flux}). This negative-polarity region is surrounded by positive-polarity flux (Figure~\ref{fig:hmi_flux}(e)--(g), (k)--(m)); traditionally such flux regions are termed as `anemone' regions, although this term is usually used in description of smaller ARs \citep{2008ApJ...673.1188A}. The negative flux profile in Figure~\ref{fig:hmi_flux}a clearly reveals that the negative-polarity region (which we denote as N$_{E}$) is fast evolving. Another AR (AR 11514) was situated very close to the AR of our interest, i.e., AR 11515 (with an east-west separation of $\sim$50 arc sec along the X-axis), with the two ARs being almost aligned in the east-west direction. Furthermore, on the southwest side of AR 11515, there lies a dispersed negative magnetic field region (labeled as N$_{2}$ in Figure~\ref{fig:overview}). The NLFFF and PFSS coronal model field models (in Figures~\ref{fig:extrapolation} and \ref{fig:extrapolation_pfss}) clearly indicate a large-scale loop structure (LSL1) that connects the leading positive polarity region of AR 11515 (marked as P$_1$ and P$_2$) with its trailing part (marked as N$_{1}$). However, a more intriguing feature is another large-scale loop structure (LSL2) that connects the positive polarity leading sub-group of AR 11515 (i.e., P$_1$ and P$_2$) to the dispersed negative polarity region marked as N$_2$.

\begin{figure}[ht!]
%\epsscale{0.}
\plotone{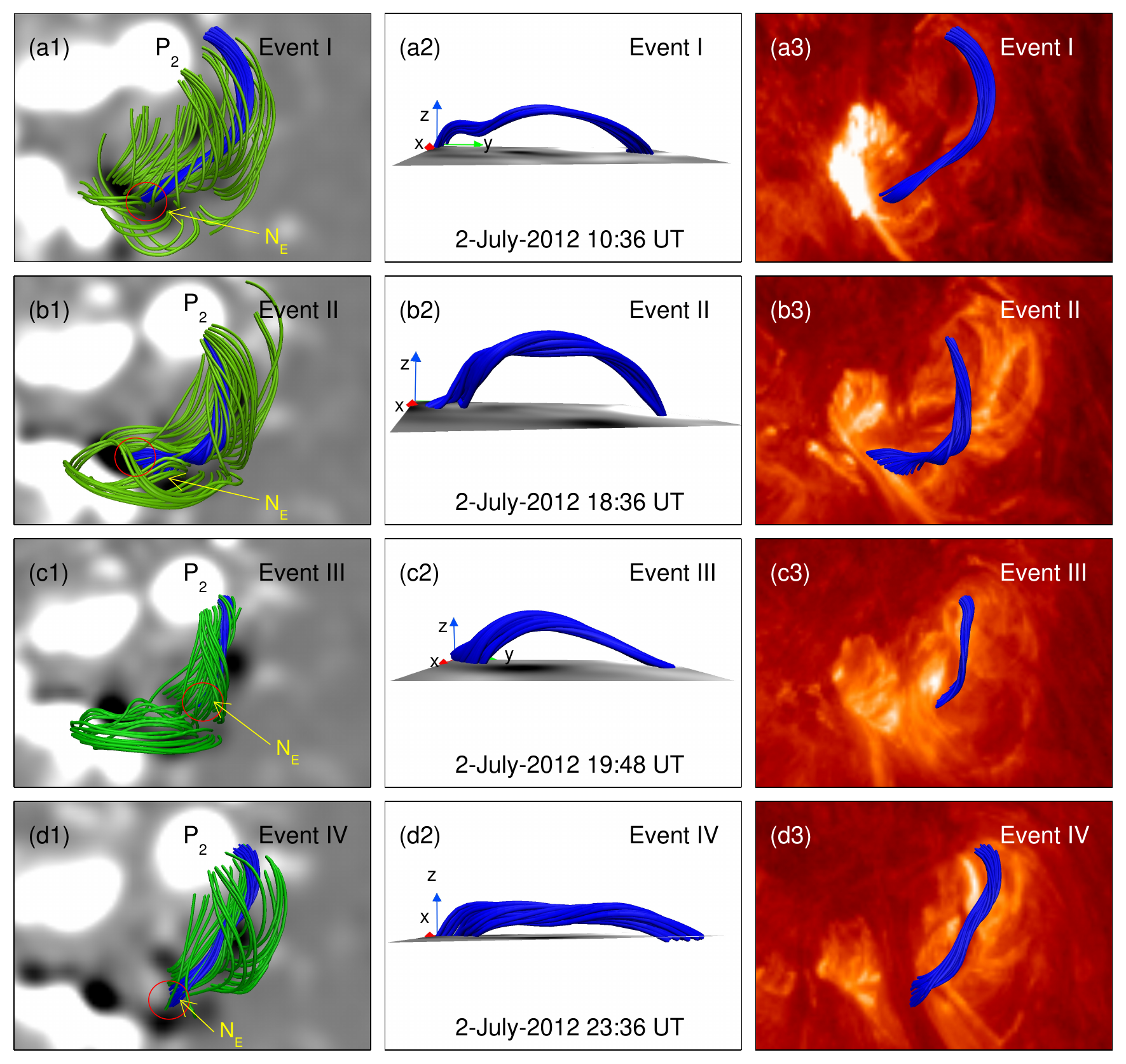}
\caption{Panels (a1)--(d1): Pre-eruptive coronal magnetic field configurations of the source region obtained from the NLFFF extrapolations using HMI vector magnetograms before events I, II, III, and IV. We show the flux rope in blue color in each panel. The source region consists of closed bipolar field lines (green), which constrain the underlying flux rope. The flux rope are formed between emerging negative flux (N$_E$) and positive polarity flux (P$_2$) in the leading part of the AR. The red circles mark the southeastern footpoint location of the flux rope in each panel, which is rooted in the rapidly changing N$_E$ region. Panels (a2)--(d2): The flux ropes are shown from side views before events I, II, III, and IV. Panels (a3)--(d3): An AIA 304 \AA~image before the respective event is plotted in the background of the flux ropes.} 
\label{fig:extrapolation_fr}
\end{figure}

A close look of the source region gives insight into the triggering process of the homologous eruptions. The compact flaring region lies close to the base of LSL1, which appears in both coronal magnetic field models, i.e., PFSS (Figure~\ref{fig:extrapolation_pfss}) and NLFFF (Figure~\ref{fig:extrapolation}). Notably, a compact flux rope (blue structure in Figures~\ref{fig:extrapolation} and \ref{fig:extrapolation_fr}) is enveloped in a system of low-lying magnetic loops (green field lines in Figure~\ref{fig:extrapolation_fr}). The southern end of the flux rope terminates at negative magnetic polarity, which is also the main triggering location of the eruptions (Figure~\ref{fig:extrapolation_fr}). We observe localized and intense brightening just before the onset of the impulsive phase of the flare at this location during all four eruptions (see region marked by the red circle in selected magnetograms and NLFFF results in Figures~\ref{fig:hmi_flux} and \ref{fig:extrapolation_fr}). Hence, we reasonably identify these early localized brightenings as signatures of tether-cutting reconnection \citep{1992LNP...399...69M, 2001ApJ...552..833M} that destabilize the southern end of the MFR. Our investigation of the photospheric magnetic field configuration during the interval of $\approx$20 hours immediately reveals the root cause of the repetitive tether-cutting reconnection and ensuing homologous eruptions: flux cancellation at and in the vicinity of the southern leg of the MFR \citep{2010A&A...521A..49S, 2011A&A...526A...2G, 2012ApJ...759..105S}. Figure~\ref{fig:hmi_flux} implies that photospheric flux near the triggering location (i.e., the southern leg of MFR) indeed presents a quite dynamic evolution, both at spatial and temporal scales. From this overall behavior of magnetic flux over the time span of homologous activity, i.e., emergence and cancellation occurring simultaneously, both processes plausibly could contribute to the reformation of the MFR before each eruption: flux cancellation could contribute to the continued strengthening of the MFR \citep{2010A&A...521A..49S, 2011A&A...526A...2G, 2014ApJ...797L..15C}, and flux emergence could contribute to the further rise of the MFR magnetic field from below the photosphere \citep{2007ApJ...669.1359S, 2007SoPh..241...99M, 2011ApJ...731L...3S}.

\begin{figure}[ht!]
\plotone{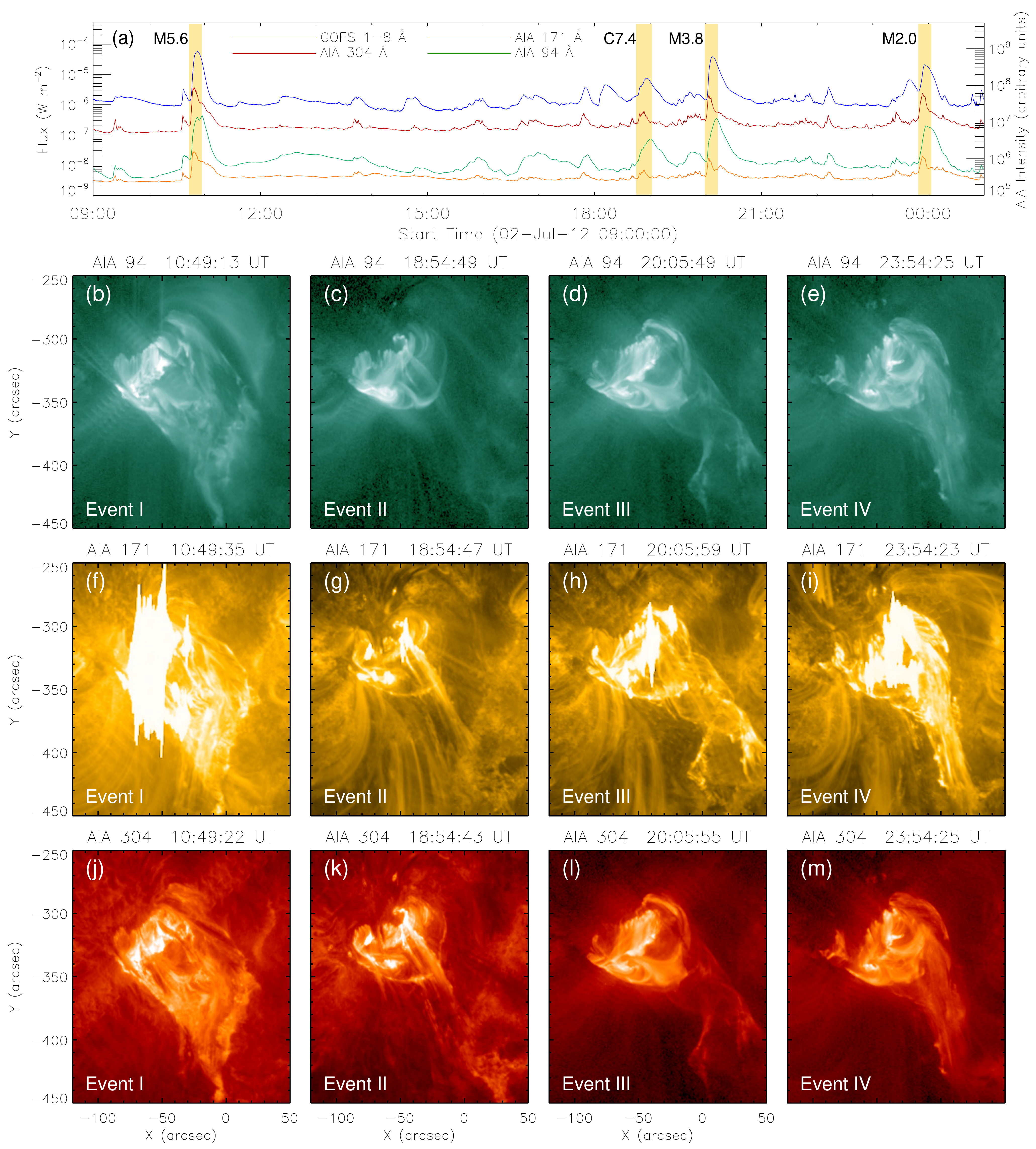}
\caption{Panel (a): Lightcurves for GOES 1--8 \AA~(blue curve), AIA 304 \AA~(red curve), AIA 171 \AA~(orange curve), and AIA 94 \AA~(green curve) channels. Four yellow-shaded regions mark the flares under consideration. The flare classes are annotated with the shaded region (i.e., M5.6, C7.4, M3.8, M2.0), corresponding to events I, II, III, and IV, respectively. Panel (b)--(d): AIA 94 \AA~images of the flaring region near the peak time of the flares. Panels (f)--(i) and panels (j)--(m) show AIA 171 and 304 \AA~images of the flaring region that are cotemporal with the AIA 94 \AA~images in (b)--(d), respectively. }
 \label{fig:goes_ew_overview}
\end{figure}

\begin{figure}[ht!]
\plotone{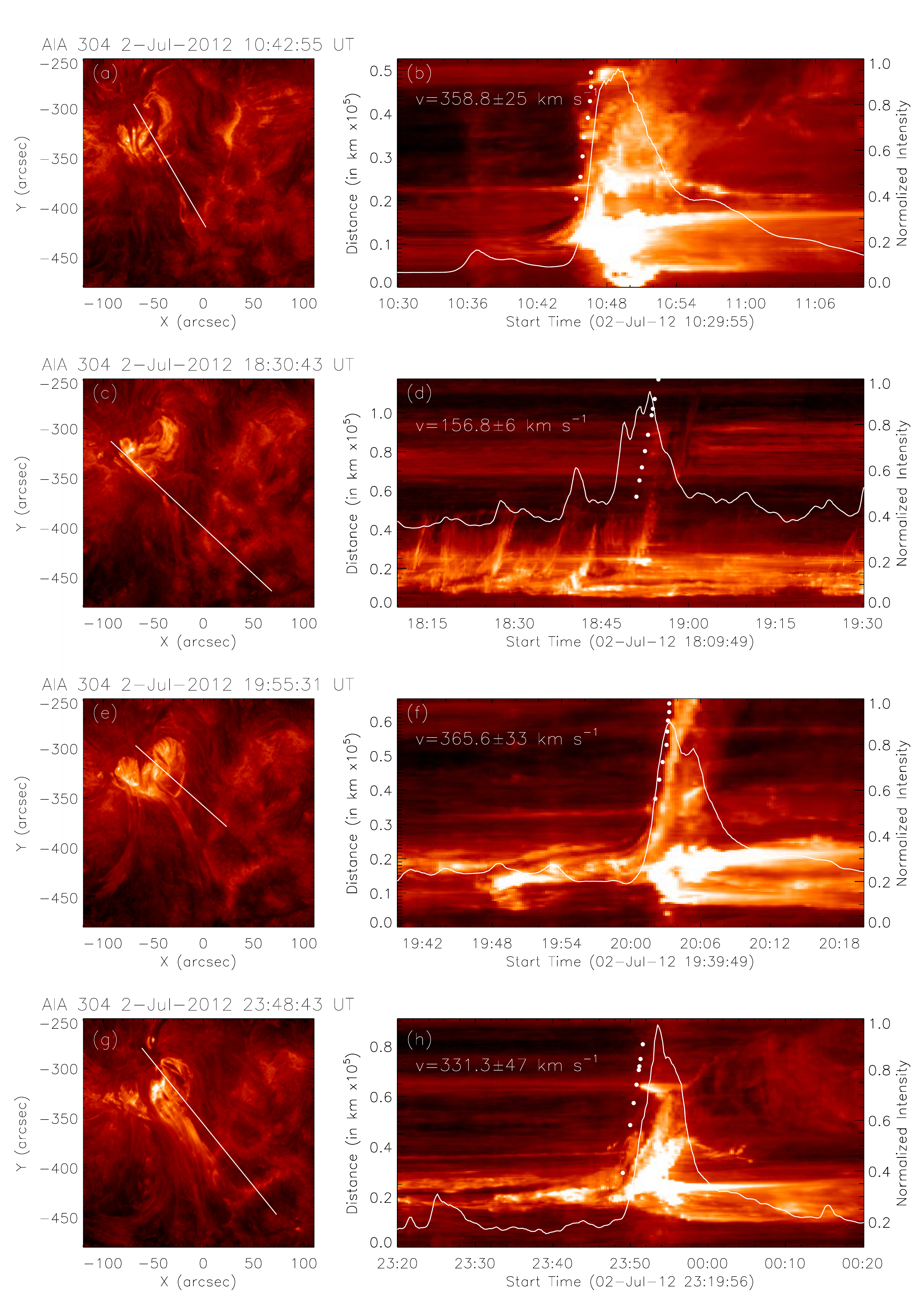}
\caption{Time-slice diagrams of the erupting filaments in AIA 304 \AA~channel. Panels (a)--(b), (c)--(d), (e)--(f), and (g)--(h) corresponds to event I, II, III, and IV, respectively. The left column shows the slices of the AIA 304 \AA~channel images, for the distance-time profiles of the eruptions. The right column shows the distance-time profiles. The speeds of the erupting structures are annotated in these panels, with the corresponding uncertainties in the measurements (i.e., 1$\sigma$ error). The flare intensity obtained from the AIA 304 \AA~channel is also overplotted with the white curves. In panels (b), (d), (f), and (h), white dots trace the leading edge of the erupting plasma. 
\label{fig:ht_merge}}
\end{figure}

In these homologous events, the erupting filament and accompanying erupting plasma from the flaring region was mostly collimated, in contrast to the cases of wider-spread flares. The collimated eruption propagated mainly in the projected southwest direction (See Figure~\ref{fig:goes_ew_overview}) and resembles a large-scale jet \citep{2016ApJ...822L..23P, 2017ApJ...851...29J, 2018MNRAS.476.1286J}. Except for event II, the fast eruptive motion of the filament is in the impulsive phase of the flare \citep{2001ApJ...559..452Z, 2005ApJ...630.1148S}, a typical feature well accepted for eruptive flares \citep{2012ASSP...33...29J, 2018SSRv..214...46G}. For event II, we observe relatively less prominent, multiple jet-like features during the pre-flare phase, while the traced part of the main filament eruption started during the flare's peak and continued after the flare's EUV peak (Figure~\ref{fig:ht_merge}b). Our observations of event II suggest that the multiple jets during the pre-flare phase are standard jets \citep{2010ApJ...720..757M}, i.e., each is driven by a confined eruption of filament/MFR, not a blowout eruption of the filament/MFR. Notably, the wide jet spire driven by an ejective filament eruption is characteristic of blowout jets \citep{2010ApJ...720..757M, 2022FrASS...920183S, 2015Natur.523..437S, 2022ApJ...927..127S}.

%show that the preceding multiple jets during the pre-flare phase resemble the standard jets \citep{2010ApJ...720..757M} i.e., the jet events do not involve filament/MFR in the interchange reconnection. Notably,} the wide jet spire driven by an ejective filament eruption is characteristic of blowout jets \citep{2010ApJ...720..757M, 2022FrASS...920183S, 2015Natur.523..437S, 2022ApJ...927..127S}.

\begin{figure}[ht!]
\plotone{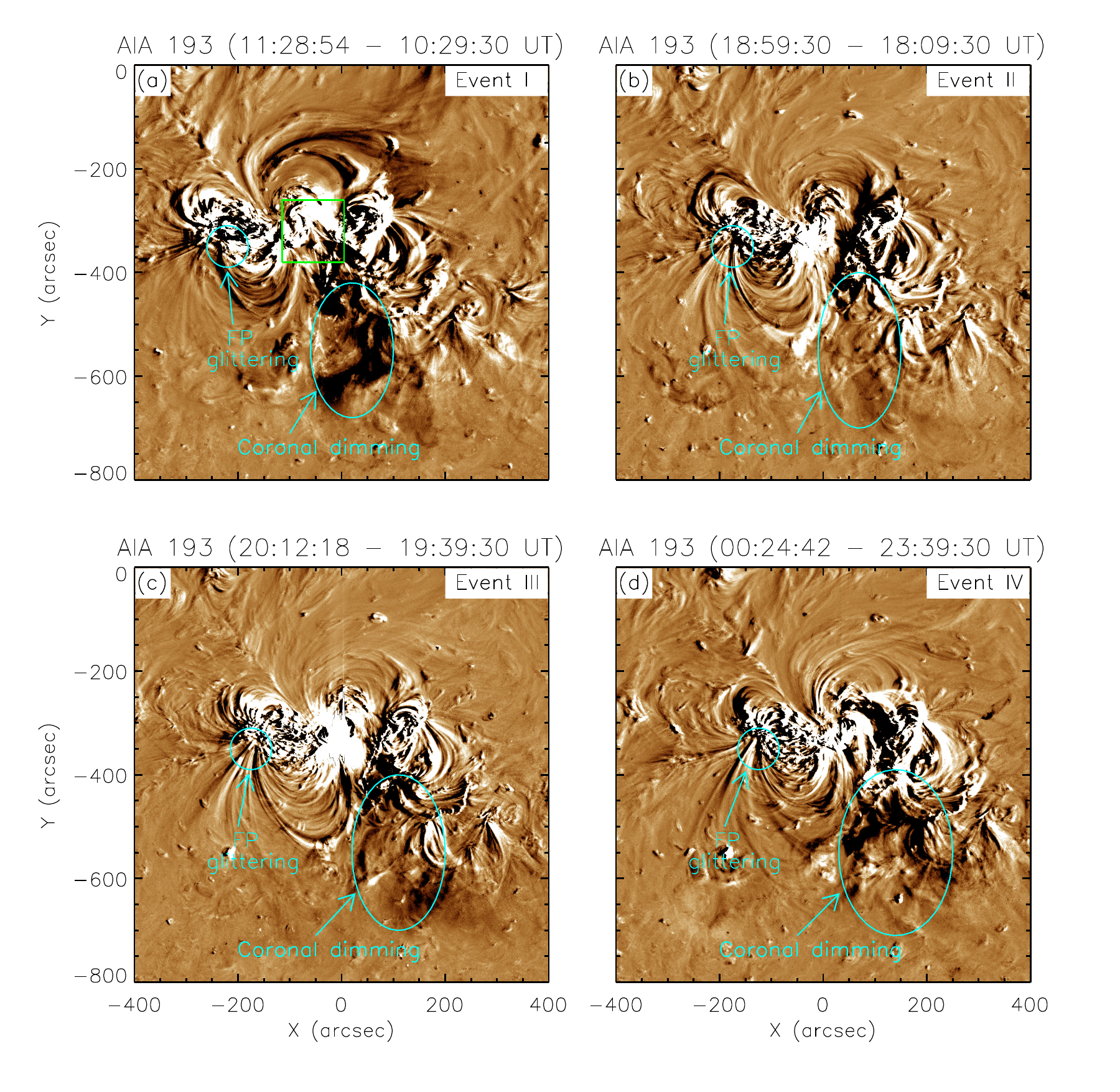}
\caption{Representative AIA 193 \AA~fixed-difference images showing the distant coronal dimming produced by the large-scale eruptions of event I (panel a), II (panel b), III (panel c), and IV (panel d). These images are created by subtracting a pre-flare image for each event. The dimming area is enclosed by the blue ellipses in all panels. The brightening and darkening at the trailing part of the AR 11515, which we call ``glittering" at the footpoints (FP), is enclosed by with a cyan circle.} 
\label{fig:dimming_mosaic}
\end{figure}

\begin{figure}[ht!]
%\epsscale{0.8}
\plotone{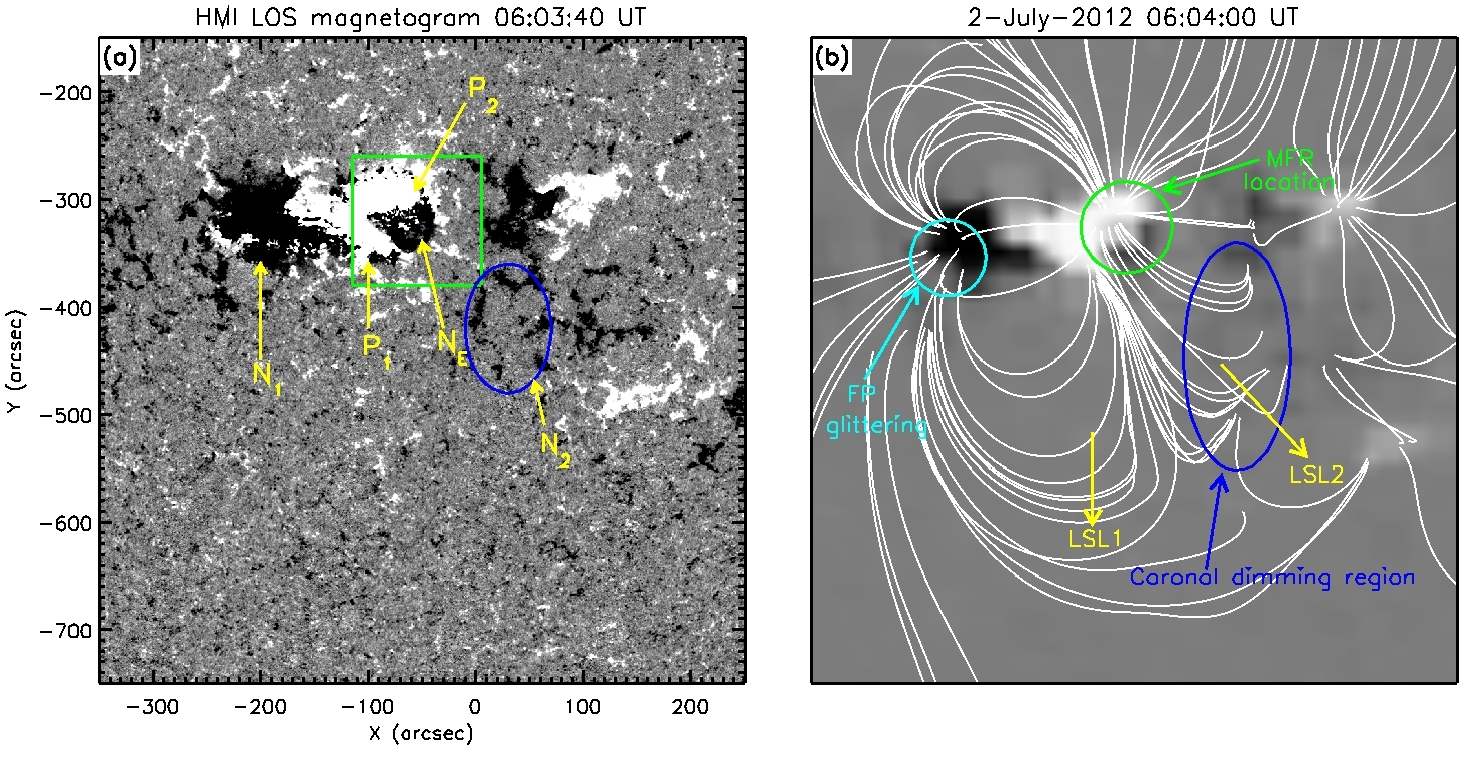}
\caption{Panel (a): HMI LOS magnetogram observed prior to event I. We point to different magnetic polarities near the filaments with arrows. We mark the flaring region with the green box. The blue-colored ellipse shows the distant negative polarity region west of the AR 11515. Panel (b): Results from the PFSS extrapolation model co-temporal with the HMI magnetogram (shown in panel a). The connectivity between the leading and trailing parts within AR 11515 (LSL1), and between positive polarities of AR 11515 and negative polarities that are in a nearby AR and at more remote locations (LSL2; P$_1$--N$_2$ and P$_2$--N$_2$). The southeastern footpoints of LSL1 showing brightening and dimming (glittering) in Figure~\ref{fig:dimming_mosaic} are inside the cyan circle. The location of the MFR obtained in the NLFFF extrapolation model is along the polarity inversion line inside the green circle (see Figure~\ref{fig:extrapolation_fr}). The blue ellipse is co-spatial with the dimming region marked by the blue ellipse in Figure~\ref{fig:dimming_mosaic}a.
\label{fig:extrapolation_pfss}}
\end{figure}

The AIA 193 \AA~difference images show that coronal dimming formed to the southwest of the flaring site following the eruption of the flux rope. This dimming was dispersed (Figure~\ref{fig:dimming_mosaic}), and occurred in the southern foot of LSL2. We also notice small-scale transient brightenings and darkenings at the eastern footpoint of LSL1, which we term footpoint glittering (see location marked as FP glittering in Figure~\ref{fig:dimming_mosaic}). Thus, we conclude that the wide CMEs result from the involvement of LSL1 and LSL2 in the process of the compact flux rope eruptions. The main source of the CME mass is the coronal-dimming location, which is disjoint from the MFR location.

\begin{figure}[ht!]
\epsscale{0.7}
\plotone{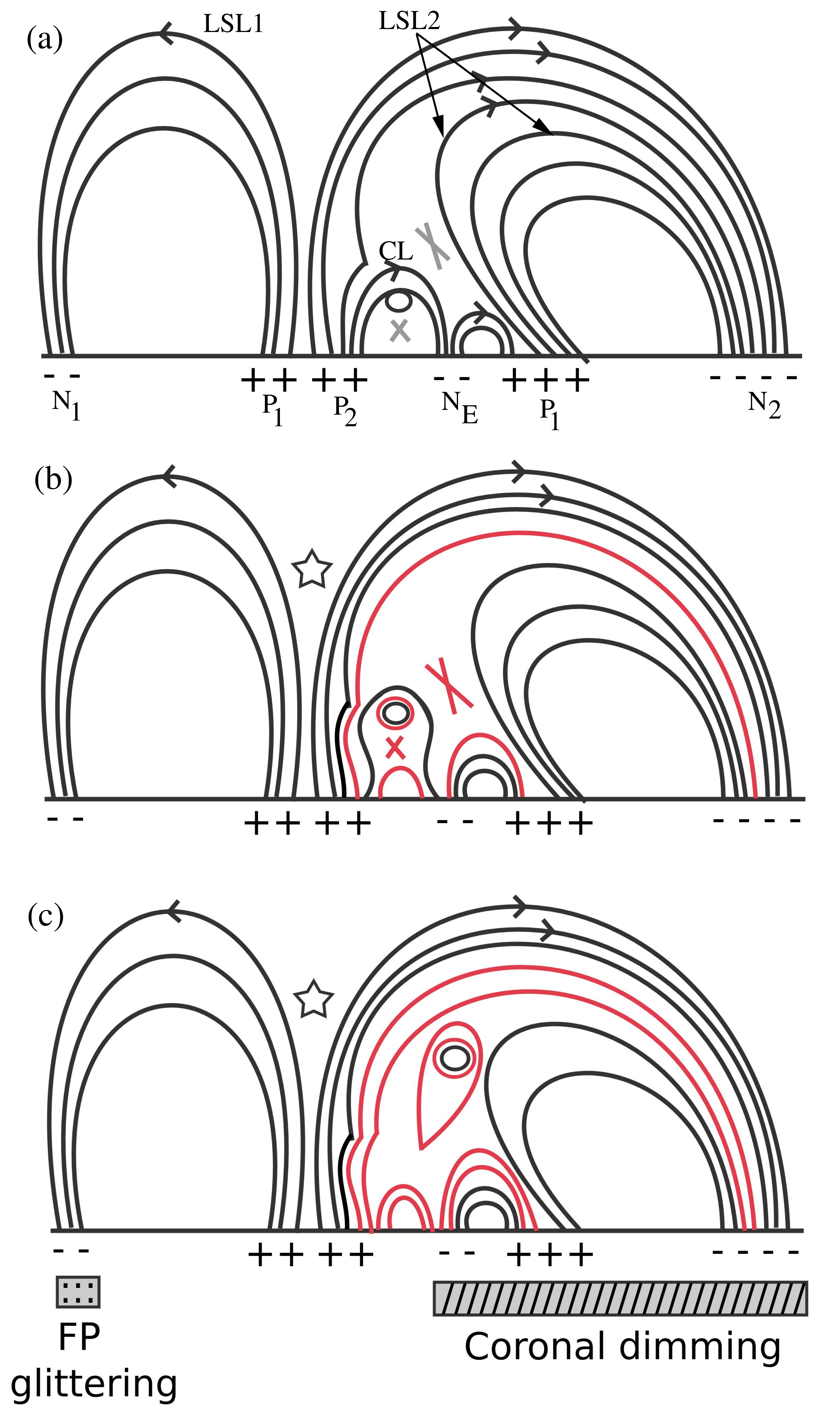}
\caption{Schematic representation for the formation of broad CMEs from homologous blowout-jet-making filament eruptions. Panel (a): the large-scale loops (LSL1) connect the N$_1$ and P$_1$ fluxes of the AR. On the right of P$_1$, we show P$_2$, next to an emerging negative polarity region (N$_E$). We show the flux rope constrained by the overlying compact loops (CL). On the right of CL, we show smaller loops connecting N$_E$ and nearby P$_1$ flux. On the right side of these small loops, we show the connectivity of P$_1$ flux with the distant negative polarity region to the west (N$_2$; LSL2 loops, see Figure~\ref{fig:overview}(d),~\ref{fig:hinode_xrt}, and \ref{fig:extrapolation_pfss}(b)). Gray X signs mark the possible reconnection sites. Panel (b): the reconnection below the flux rope (described as ``tether-cutting reconnection" and ``internal reconnection" in the text) destabilizes the system, and the reconnection between the overlying loops LSL2 (called "external reconnection" in the text), reconnects LSL2 loops, and creates a pathway for the expulsion of the flux rope. We show the reconnected field lines in red. In panels (b) and (c), the stars between LSL1 and LSL2 shows the possible site of quasi-parallel reconnection. Panel (c): the external reconnection between CL and LSL2 produced two sets of field lines, one above the LSL2 and the other above the small loops. The brightening and dimming were observed near the eastern footpoints of LSL1, which we label as footpoint (FP) glittering. The extent of the dimming resulting from the blowout eruption of jets and overlying loops is also indicated in panel (c). 
\label{fig:schematic}}
\end{figure}

In Figure~\ref{fig:schematic}, we provide a schematic representation to explain the formation of the wide CME from each of our blowout-jet-making filament eruptions. In Figure~\ref{fig:schematic}a, we sketch the configuration of large-scale magnetic fields in relation to the compact MFR prior to the eruption. The multi-wavelength imaging in various EUV channels and in H$\alpha$ together with coronal magnetic field modeling converge to this scheme: (1) A compact MFR (filament) resides at the foot of large-scale magnetic field that is rooted in the positive-polarity flux surrounding the flux rope (region P$_2$). (2) Above the MFR, the large-scale field mostly diverges to two prominent locations -- the trailing part of AR 11515, N$_1$, and the distant negative polarity region, N$_2$. (3) From points (1) and (2), we infer that the opposite ends of the compact MFR lie at positive polarity region P$_2$ and at the rapidly evolving negative flux region N$_E$.

The subsequent development of erupting MFR and its interaction with large-scale loops (LSL2) produce the observed wide CME. Panels (b) and (c) show the subsequent evolution as the compact the MFR becomes destabilized and erupts, forming any of the four homologous eruptions. The MFR becomes activated due to tether-cutting reconnection involving the MFR field and low-lying loops in the small magnetic arcade linking N$_E$ and P$_2$. The MFR eruption drives both ``internal'' and ``external'' reconnection. The internal reconnection ensues between the legs of the enveloping field lines stretched by the erupting flux rope, which further unleashes and accelerates its upward expansion. Subsequently, the upward-accelerated MFR drives external reconnection between the core fields and LSL2. The two reconnection sites are marked by cross signs (Figures \ref{fig:schematic}(a)--(b)), and post-reconnection field lines are drawn in red in Figures~\ref{fig:schematic}(b)--(c). The external reconnection alters the LSL2 field lines and creates a ``pathway'' for the ensuing outward expulsion of the flux rope. Thus, the progression of the flux rope eruption follows the trajectory depicted by the large LSL2 loops. The process continues with the sideways, jet-like escape of the MFR by sequentially blowing out the loops in the large LSL2 arcade. The eventual eruption of the flux rope therefore leads to strong dimming region (indicated in Figure~\ref{fig:schematic}(c)) extending from the AR up to the N$_2$ which essentially forms the source region of the associated wide CME. This is similar to the process whereby jets in the leg of a larger loop blew out that larger loop in \cite{2016ApJ...822L..23P}. Contextually, \cite{2015ApJ...813..115L} presented a case study employing multi-point and multi-wavelength observations of a large blowout jet from an ``anemone" active region, which showed that the CME was not the extension of the jet in the coronagraphs, but was triggered by the jet event with jet becoming its core.

%Through the gradual advancement of this process, the flux rope ultimately blows out the prominent LSL2 loops, thereby causing the expansion of the pronounced dimming region (depicted in Figure~\ref{fig:schematic}(c)) spanning from from the AR up to the N$_2$.} 

%The external reconnection alters the LSL2 field lines and creates a ``pathway'' for the ensuing outward expulsion of the flux rope. The eruption of the flux rope continues along the curve of the large LSL2 loops. As this process continues, the flux rope eventually blows out the large LSL2 loops, making the strong dimming region (indicated in Figure~\ref{fig:schematic}(c)) extend from the AR up to the N$_2$.

Another less prominent but important process is the interaction between the expanding LSL2 loops with the adjacent large LSL1 loops, which also drives reconnection in the corona. Here we note that the LSL1 and LSL2 loops are quasi-parallel. As noted in an earlier study, the reconnection between quasi-parallel loops leads to weak flaring and a consequent faint signature of energy release \citep{2019ApJ...871..165J}. The signature of interaction between LSL1 and LSL2 is seen at the far footpoint of LSL1 rooted in N$_1$, where the EUV difference images show subtle brightening and dimming, termed as footpoint (FP) glittering (Figure~\ref{fig:schematic}(c), see also Figures~\ref{fig:dimming_mosaic} and \ref{fig:extrapolation_pfss}). Furthermore, we suggest that the rapidly evolving magnetic flux in N$_E$ (see Figure~\ref{fig:hmi_flux}) ensures the rebuilding of the MFR before each compact blowout-jet explosion, and also plausibly generates the circumstances resulting in repeated tether-cutting that triggers the jet-flare events by destabilizing the MFR. Thus, we find that the CME in our homologous eruptions does not acquire most of its mass and dynamics directly from the compact MFR eruption, as happens in traditional eruptive flares, but results from complex external reconnections. This also provides support for a plausible scenario of the development of large-scale CME structure triggered by compact blowout jet.

\begin{acknowledgments}
We would like to thank the SDO and SOHO teams for their open data policy. SDO (2010--present) is NASA's mission under the Living With a Star (LWS) program. SOHO (1995--present) is a joint project of international cooperation between the ESA and NASA. We gratefully acknowledge the LASCO CME catalog, generated and maintained at the CDAW Data Center by NASA and The Catholic University of America in cooperation with the NRL. Hinode is a Japanese mission developed and launched by ISAS/JAXA, with NAOJ as domestic partner and NASA and STFC (UK) as international partners. It is operated by these agencies in co-operation with ESA and NSC (Norway). We further acknowledge use of the GONG data from NSO, which is operated by AURA under a cooperative agreement with NSF and with additional financial support from NOAA, NASA, and USAF. A.C.S. and R.L.M. received funding from the Heliophysics Division of NASA's Science Mission Directorate through the Heliophysics Supporting Research (HSR; grant No. 20-HSR20 2-0124) Program, and through the Heliophysics System Observatory Connect (HSOC; grant No. 80NSSC20K1285) Program, and A.C.S. received additional support through the MSFC Hinode Project. We are thankful to Dr. Thomas Wiegelmann for providing the NLFFF code. The IDL based code utilized for the computation of twist number is available online at \url{http://staff.ustc.edu.cn/\~rliu/qfactor.html}. We thank Dr. Prabir K. Mitra for helpful discussions on the extrapolation techniques. We are thankful to the anonymous reviewer of the paper for providing constructive comments and suggestions that have significantly enhanced the quality of the paper.\\ 
\textit{Software:} SolarSoftware \citep{2012ascl.soft08013F}, VAPOR \citep{2007NJPh....9..301C}. 
\end{acknowledgments}

\vspace{5mm}
%\newpage
\bibliography{ms_v_mar2023}{}
\bibliographystyle{aasjournal}

\begin{table}
\begin{center}
\caption{The four homologous CMEs and their joint flares produced by AR 11515 on 2012 July 2.}
\begin{tabular}{cccccccccc}
	\hline
	Event Number & Date  & \multicolumn{3}{c}{CME parameters}&\multicolumn{5}{c}{Flare parameters}\\
	 \cline{3-4} \cline{5-10}
	 & &Time\footnote{CME time corresponds to the first appearance time in the LASCO C2 FOV}& CME angular &  CME Speed\footnote{CME speed is obtained from the CDAW CME catalog. This CME speed is the average speed calculated from LASCO C2 and C3 observations.}  & Start & End & Peak & Heliographic & Class \\
	& & (UT) &width ($\arcdeg$) & (km~s$^{-1}$) & (UT) & (UT) & (UT) & Coordinate&\\
	\hline

I &2012/07/02&	11:24:04 &		125 &	313  &   10:43 &10:57 &10:52 &S22E02                        & M 5.6   \\
II&2012/07/02 &	19:36:04 &	96 &	503 &	  18:45 &19:02& 18:56& S17W01                        & C 7.4    \\ 
III&2012/07/02 &20:24:05 &	145 &		527 	&      19:59& 20:13& 20:07& S17W01    &                     M 3.8    \\ 
IV& 2012/07/03 &	00:48:04& 		97 &	400 &	  23:49 &00:03 &23:56& S17W03  & M 2.0  \\ 
\hline
\end{tabular}
\label{tab:hom_cme}
\end{center}
\end{table}

%\textbf{The accompanying animation shows the evolution of eruptive flares in AIA 94 \AA, AIA 171 \AA, and AIA 304 \AA~channels along with their corresponding light curves from 10:30 UT on 2012 July 2 to 00:20 UT on 2012 July 3. The total duration of the animation is 17 s.\\(An Animation associated with this figure is available.)}

%\textbf{The accompanying animation presents AIA 193 \AA~fixed-difference images, showing the evolution of coronal dimming, along with the GOES SXR light curves from 10:38 UT on 2012 July 2 to 00:59 UT on 2012 July 3. The total duration of the animation is 24 s.\\(An Animation associated with this figure is available.)}}

%\textbf{The accompanying animation shows a time-evolution of the HMI continuum and LOS magnetogram images from 04:56 UT on 2012 July 2 to 00:56 UT on 2012 July 3. The total duration of the animation is 20 s.\\(An Animation associated with this figure is available.)}}

%Panels (a)--(d): Pre-eruptive coronal magnetic field configurations of the source region obtained from the NLFFF extrapolations using HMI vector magnetograms before events I, II, III, and IV. We show the flux rope in blue color in each panel. The source region consists of closed bipolar field lines (green), which constrain the underlying flux rope. The flux rope are formed between emerging negative flux (N$_E$) and positive polarity flux (P$_2$) in the leading part of the AR. The red circles mark the southeastern footpoint location of the flux rope in each panel, which is rooted in the rapidly changing N$_E$ region.

\end{document}